# Identification and time-resolved study of YIG spin wave modes in a MW cavity in strong coupling regime

Angelo Leo[1,2,*], Anna Grazia Monteduro[1,2], Silvia Rizzato[2], Luigi Martina[1,3], Giuseppe Maruccio[1,2,*]

[1]*Department of Mathematics and Physics, University of Salento, Via per Arnesano, 73100, Lecce, Italy*
[2]*CNR NANOTEC - Istituto di Nanotecnologia, Via per Arnesano, 73100 Lecce, Italy,*
[3] *INFN – Sezione di Lecce, Via per Arnesano, 73100 Lecce, Italy*

**ABSTRACT**

Recently, the hybridization of microwave-frequency cavity modes with collective spin excitations attracted large interest for the implementation of quantum computation protocols, which exploit the transfer of information among these two physical systems. Here, we investigate the interaction among the magnetization precession modes of a small YIG sphere and the MW electromagnetic modes, resonating in a tridimensional aluminum cavity. In the strong coupling regime, anti-crossing features were observed in correspondence of various magnetostatic modes, which were excited in a magnetically saturated sample.

Time-resolved studies show evidence of Rabi oscillations, demonstrating coherent exchange of energy among photons and magnons modes. To facilitate the analysis of the standing spin-wave patterns, we propose here a new procedure, based on the introduction of a novel functional variable. The resulting easier identification of magnetostatic modes can be exploited to investigate, control and compare many-levels hybrid systems in cavity- and opto-magnonics research.

* corresponding authors: angelo.leo@unisalento.it, giuseppe.maruccio@unisalento.it

*INTRODUCTION*

Combining different fundamental excitations is a recent route for quantum computation applications, with the promise to stimulate the development of new hybrid quantum technologies and protocols. Indeed, it was suggested that encoding information in different physical systems can provide advantages in overcoming the strict requirements in terms of decoherence timescales and capacity to process the information, which can be difficult to match together. In this respect, a crucial requirement is the achievement of a strong coupling regime between the respective fundamental excitations in two physical systems. Recent findings demonstrated the capability to obtain a robust hybridization among light quanta and different excitations at low temperature, by employing trapped atoms [1], nitrogen vacancy (NV) centers in diamonds [2], superconducting (SC) qubits [3] and spin impurities in Si [4]. In this frame, magnons exhibited strong stability in coupling with photons, when they are excited in ferro/ferri-magnetic (FM) materials, especially if Yttrium Iron Garnet (YIG) single crystals are used [5]. In contrast with paramagnetic spin ensembles, which at room temperature (RT) are weakly coupled to the photons, YIG presents at least a three orders greater net spin density, which permits to get the strong coupling. To couple spin waves (SW) with electromagnetic (EM) signals, a convenient way is to confine the YIG in a three-dimensional (3D) microwave (MW) cavity [6]. It was reported that this yields the stable formation of two-level systems and magnon-cavity polaritons, as a consequence of the hybridization among MW photons and the fundamental magnetostatic mode (also known as Kittel mode and corresponding to ferromagnetic resonance (FMR)) [7-13]. Even non-uniform magnetostatic modes (MSMs) can be sustained by the material depending on its shape and they can be also coupled to cavity modes [14, 15]. The resulting pattern of spectral features is more complex and its association with specific modes can be not straightforward.

In the present work, we investigate the interaction among the magnetization precession modes, in a small magnetically-saturated YIG sphere, and the MW electromagnetic modes, resonating in a tridimensional aluminum cavity, at room temperature. A rich spectrum characterized by several anti-crossing features is observed, because of the strong coupling regime in correspondence of various magnetostatic modes. Time-resolved studies show evidence of Rabi oscillations, demonstrating (for the first time at room temperature) coherent exchanges of energy among photons and the involved magnons modes. For facilitating the analysis of the stationary spin-wave patterns, here we propose a new procedure, based on the introduction of a novel functional variable, related to the magnetic characteristics of the FM material and to the applied external electromagnetic field. Notably, plotting the data with respect to this variable, we obtain a direct identification of the involved MSMs.

## EXPERIMENTAL SETUP AND CAVITY MODES

The investigated system is composed by a single crystal YIG sphere, with 1 mm of diameter [16], located into an aluminum cavity with inner dimensions of $44 \times 22 \times 9$ mm$^3$. The Yttrium Iron Garnet (YIG, $Y_3Fe_5O_{12}$) is chosen for its peculiar characteristics. In particular, its magnetic moment comes from $Fe^{+3}$ ions in the $6S_{5/2}$ ground state and YIG behaves as a ferrimagnetic insulator with a 550 K Curie temperature, a typical saturation magnetization $M_S = 0.178$ T, spin density of $4.22 \cdot 10^{27}$ m$^{-3}$, an exchange constant $\alpha = 3 \times 10^{-12}$ cm$^2$ [17]. A crucial characteristics for the applications is the low magnetic damping and a correlated narrow linewidth of $2.3 \cdot 10^{-3}$ mT [17]. This property has favored the use of the YIG crystals in optical and radiofrequency devices, such as microwave oscillators, circulators and optical isolators, since many decades.

First, the YIG sphere is placed in a central position at the bottom of one semi cavity of the ground wall as shown in **Fig. 1.a.** Such a position corresponds to the magnetic antinode for the fundamental (TE$_{101}$) mode [8], in order to maximize the interaction of magnonic modes with the MW field (**Fig. 1.b**). Then, the cavity resonator is placed between the poles of a GMW electromagnet (**Fig. 1.e**) generating a magnetostatic field, whose intensity is swept from 250 mT up to 330 mT (with steps of 0.2 mT). Similarly, for further studies on the second (TE$_{102}$) mode, the YIG sphere is placed on one of the three TE$_{102}$ magnetic antinodes, located on the junction plane of the two semi cavities (this configuration is reported in **Fig. 1.c**). More precisely, the sphere is put close to the rounded wall of the resonator, on one of the lateral antinodes as shown in **Fig 1.d**. In this case, the magnetostatic field intensity is swept from 360 mT to 440 mT, in order to get the strong coupling regime. The resonator is excited by an Agilent MXG N5183A signal generator, while transmission measurements are performed by an Agilent MXA N9010 spectrum analyser. Both devices are controlled by a homemade LABVIEW software. Specifically, the frequency is swept in a range of 320 MHz, centered around the first (or the second) cavity eigenfrequency, at fixed magnetic field, and spectroscopic measurements were performed applying a 0 dBm input power.

At room temperature, zero DC magnetic field and loaded with the YIG sphere, experimentally the cavity exhibits the TE$_{101}$ mode at $\omega_c/2\pi = 8.401$ GHz, the TE$_{102}$ mode at $\omega_c/2\pi = 10.361$ GHz. The loaded quality factor $Q_L$ of the mere cavity with YIG at TE$_{101}$ is 4000, with insertion loss $IL = -33.1$ dB. This leads to an estimated intrinsic Q-factor $Q_i = Q_L/(1 - 10^{IL/20}) \simeq 4100$. The second mode exhibits a $Q_L$ of 4300, with $IL = -30.16$ dB and $Q_i = 4450$. In these conditions, the system remains lossy coupled to the measurement setup. At both the TE$_{101}$ and TE$_{102}$ magnetic antinodes, the insertion of the YIG sphere does not perturb significantly the EM signal, in absence of a drive magnetostatic field. Indeed a resonant frequency shift of less than 0.1% is observed. Furthermore, the ratio between the crystal volume $V_{YIG}$ and the magnetic modal volume $V_c$ is $V_{YIG}/V_c \approx 2 \cdot 10^{-4}$, which also justifies the observed negligible variation of the quality factor.

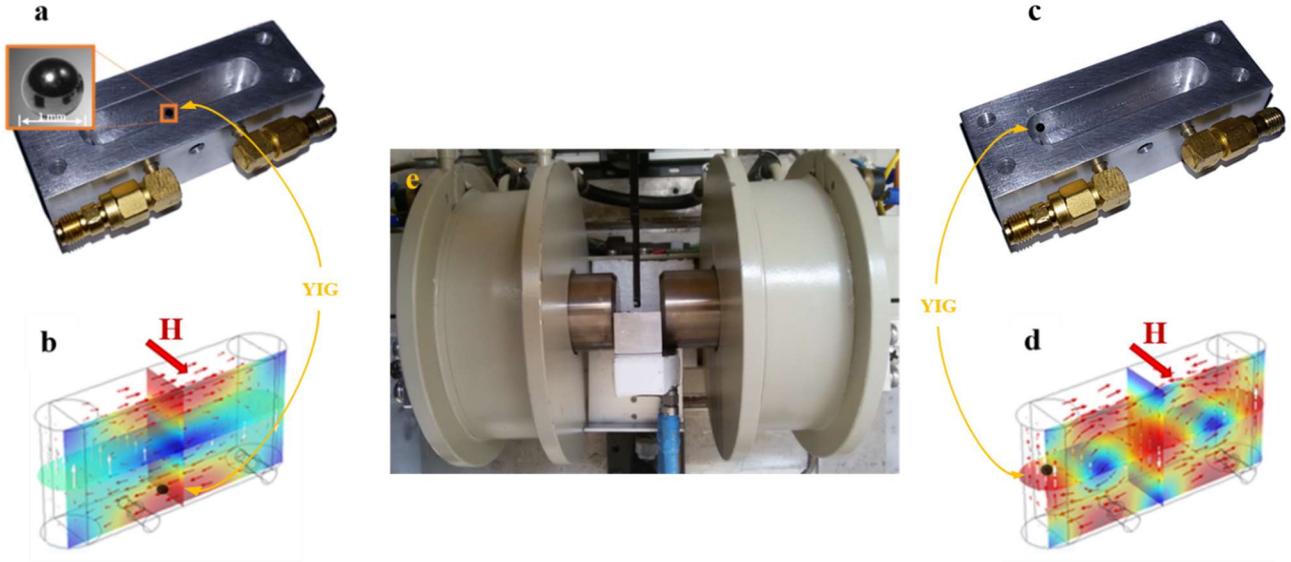

*Figure 1. (a) 3D aluminium semi-cavity loaded by a YIG sphere and (b) FEM-simulated MW magnetic field distribution for the loaded cavity at $TE_{101}$ in a perpendicular static magnetic field **H**. (c) semi-cavity loaded by YIG sphere on lateral side and (d) corresponding MW magnetic field for $TE_{102}$ (black dots on FEM simulations refer to position of sphere, in correspondence of the magnetic antinodes in the two configurations). (e) Experimental setup showing the loaded cavity between the electromagnet poles for application of the static magnetic field.*

## STRONG COUPLING REGIME

The loaded-cavity spectrum as a function of the perpendicular magnetic field is reported in **Fig. 2**. While measuring at frequencies around the first photonic mode, sweeping of the magnetostatic field intensity gives a rich magnonic spectrum, which is characterized by the presence of anti- crossings points around $TE_{101}$, as shown in the 2D map in **Fig. 2.b**. The ferromagnetic resonance (FMR, known as Kittel mode [7]) lies at 294.1 mT, as indicated by a yellow arrow. Additionally, a series of MSMs are exhibited, which are pointed out by dotted yellow lines, whose separation gradually reduces towards lower magnetostatic fields. **Fig. 2.a** illustrates how the signal amplitude at $TE_{101}$ varies as a function of the magnetic field: the transmitted electromagnetic signal takes the form of Fano resonances [18, 19] and it is significantly reduced in correspondence of the avoided crossings, with a splitting of resonance peak (in **Fig. 2.b**). On the other hand, the 2D map at $TE_{102}$ shows only two clearly visible avoided crossings (**Fig 2.c**), in addition to the main uniform precession resonance (FMR) at 391.0 mT, indicated by the yellow arrow. The relative signal amplitude from the cavity at $TE_{102}$ as a function of static magnetic field is reported in **Fig. 2.d**.

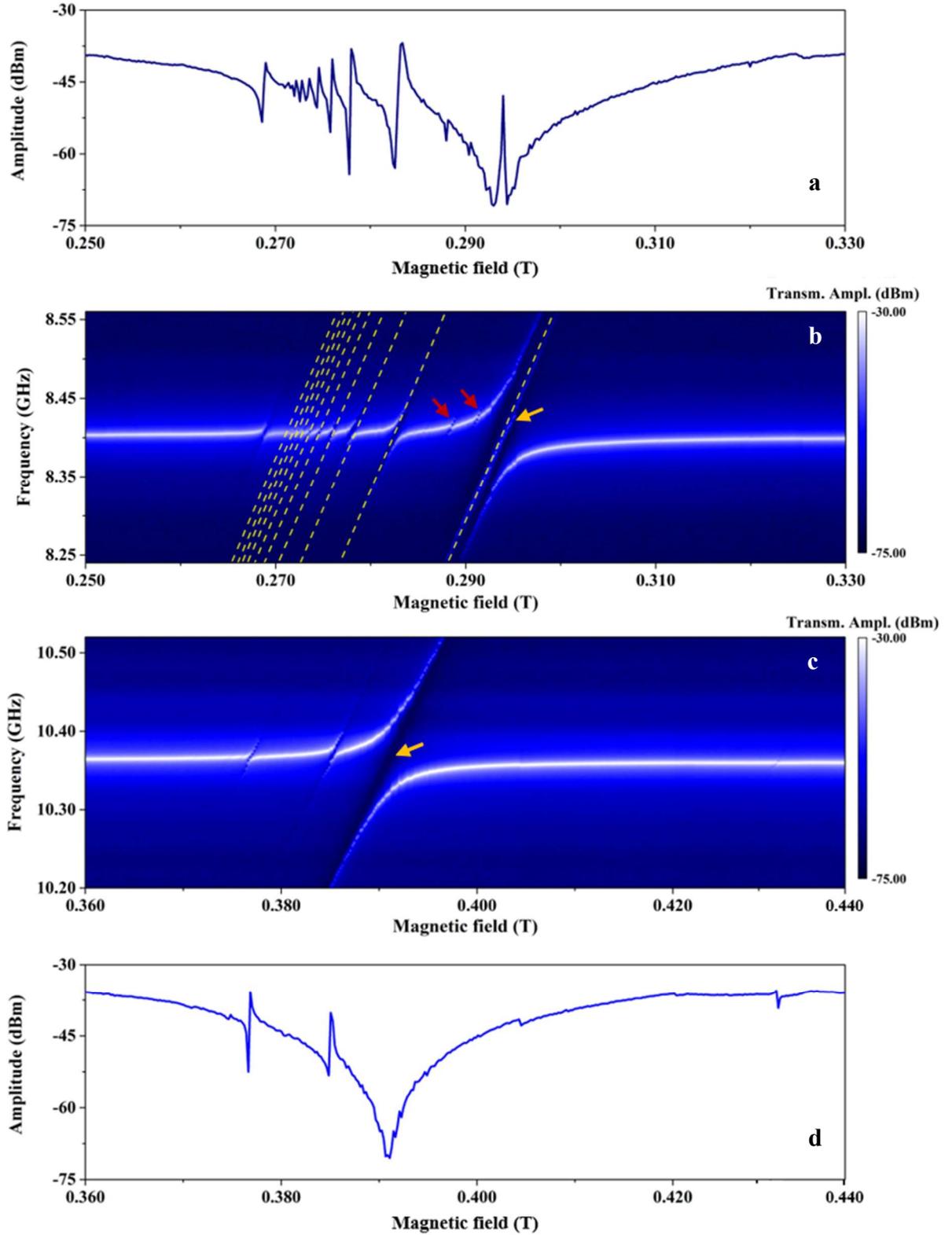

*Figure 2. (a)* Transmission amplitude at $TE_{101}$ = 8.401 GHz for magnetic field ranging from 250 mT to 330 mT. *(b)* Cavity response as a function of bias magnetic field and frequency near the fundamental $TE_{101}$ mode. Dashed yellow lines in the 2D maps of amplitude refers to identified MSMs. *(c)* Cavity response near $TE_{102}$ for magnetostatic field ranging from 360 mT to 440 mT. *(d)* Transmission amplitude at $TE_{102}$ = 10.361 GHz for magnetic field ranging from 360 mT to 440 mT (in the second configuration with sphere on cavity lateral side).

The avoided level crossings in the map of transmitted signal amplitude $T$ can be described by means of an input-output formalism [9, 13, 20, 21]:

$$T(\omega) = \frac{\kappa_c}{j(\omega-\omega_c)-\frac{1}{2}(2\kappa_c+\kappa_{int})+\sum_i \frac{|g_i|^2}{-\frac{1}{2}\gamma_i+j(\omega-\omega_i)}}, \quad (1)$$

where $\kappa/2\pi = (2\kappa_c + \kappa_{int}) = \omega_c/2\pi Q_L$ is the photonic damping for each mode resonating at $\omega_c/2\pi$, which takes into account the damping $\kappa_c$ through the single connectors and the internal damping $\kappa_{int}$ associated to the mere aluminium cavity, $j$ is the imaginary unit number, the index $i$ identifies a specific magnetostatic mode, whose frequency linewidth is $\gamma_i$ and the interaction strength of the hybrid mode for the whole magnonic system is

$$\frac{g_i}{2\pi} = \frac{g_{0,i}\sqrt{N}}{2\pi} = \frac{\left(\eta\Gamma\sqrt{\frac{(\mu_0\hbar\omega_c)}{V_c}}\right)}{2\pi}\sqrt{N}, \quad (2)$$

which is related to the coupling strength $g_{0,i}/2\pi$ for a single interacting spin through the number $N$ of net spins in the examined sample [22]. Notably, $\frac{g_i}{2\pi}$ refers to coupling strength between the two (magnonic and photonic) states and is evaluated as the frequency mismatch among the resonant peaks of the hybrid modes (**Fig. 3**, on the right). The quantity $\Gamma = g\mu_B/\hbar$ is the gyromagnetic ratio of spin ensemble, where $g$ is Landé g-factor, $\mu_B$ is Bohr magneton, $\hbar$ is reduced Planck constant and $\mu_0$ is the vacuum permeability. The spatial overlap coefficient $\eta = \int_{spher} \frac{\vec{H}\cdot\vec{M}}{H_{max}M_{max}V_{YIG}} dV$ between the two subsystems (i.e. cavity and sphere modes) is calculated taking into account the driving MW magnetic field $\vec{H}$ and the complex time-dependent off-$z$ axis sphere magnetization $\vec{M}$ for the considered mode, while $H_{max}$ and $M_{max}$ correspond to their maximum values in the sphere volume $V_{YIG}$. As a further figure, the cooperativity of the two levels system is defined as $C_i = g_i^2/\gamma_i\kappa$.

In **Fig. 3**, details of the spectra of cavity toward hybridization, near the fundamental magnestostatic mode (FMR), are reported for cavity modes TE$_{101}$ and TE$_{102}$ at fields indicated by the vertical lines of corresponding colour. When the subsystems are fully coupled, $g/2\pi$ corresponds to the frequency mismatch between the resonance peaks, and broadenings $\gamma/2\pi$ and $\kappa/2\pi$ of both waves modes are comparable. In these conditions, obtained when the magnetostatic field intensity is 294.1 mT for the TE$_{101}$ and 391.0 mT for the TE$_{102}$ mode, cavity dissipation rates $\kappa/2\pi$ are 3.2 MHz and 3.0 MHz, respectively. Finally, the magnonic damping $\gamma/2\pi$ is 2.5 MHz. For a FM sphere, the uniform precession frequency $\omega_{FMR}/2\pi$ is related to the external field $H_0$ by $\frac{\omega_{FMR}}{2\pi} = \Gamma H_0$ [23]. Since the cavity mode frequencies and $\omega_{FMR}/2\pi$ must match when the two subsystems are strongly coupled, it is then possible to estimate the gyromagnetic ratio $\Gamma$ by posing $\omega_{FMR} = \omega_c = 2\pi\Gamma H_0$ and by substituting the corresponding $H_0$ values then we obtain $\Gamma \approx 28.76$ GHz/T for the YIG sphere with respect to the value 28 GHz/T reported in [24]. In **Table 1**, the magnetostatic fields as well as the evaluated coupling parameters, photonic/magnonic dampings and cooperativity $C$ are summarized for all the observed anticrossing points (and will be discussed in more details in the next section with reference to the MSM identification).

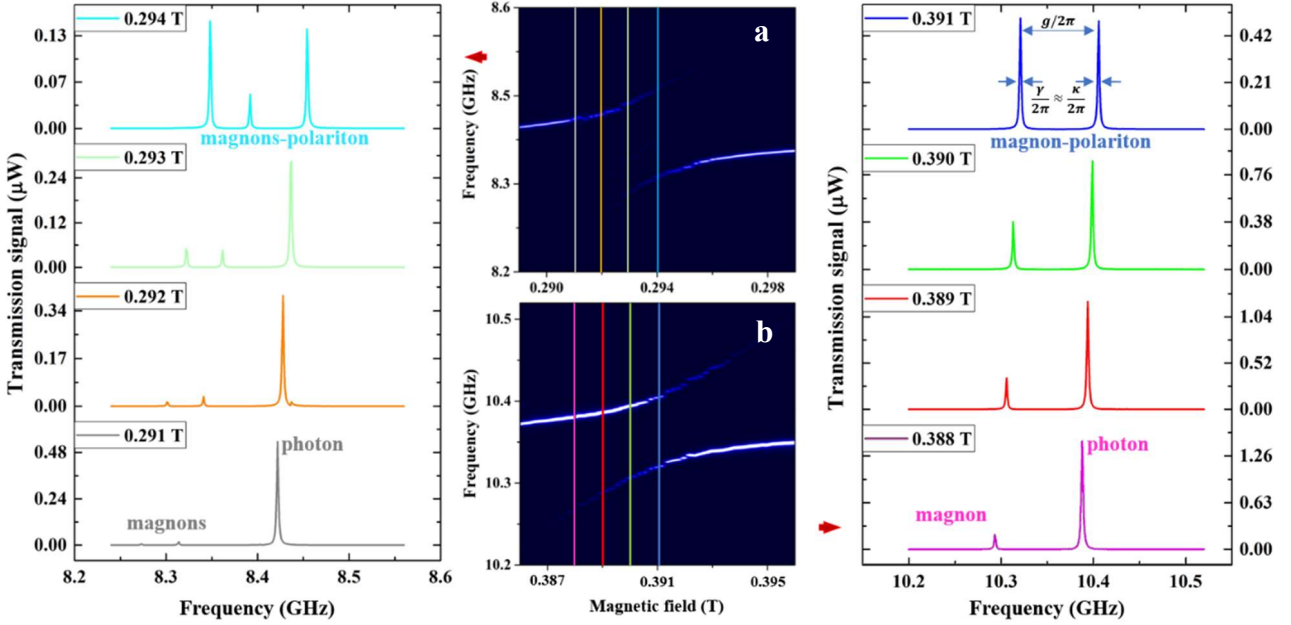

*Figure 3*. *Strong coupling between the fundamental magnetostatic mode in the YIG sphere and the $TE_{101}$ and $TE_{102}$ cavity modes.* ***(a)*** *2D maps around $TE_{101}$ when the magnetostatic field ranges between 289 mT and 299 mT. The spectra corresponding to coloured lines are reported on the left, where formation of magnons-polariton systems is shown.* ***(b)*** *2D maps around $TE_{102}$ when the magnetostatic field ranges between 386 mT and 396 mT. The spectra corresponding to coloured lines are reported on the right. When the systems are fully coupled, $g/2\pi$ is the frequency mismatch between the resonance peaks, and broadenings $\gamma/2\pi$ and $\kappa/2\pi$ of both standing waves modes are similar.*

## *IDENTIFICATION OF MSM*

In order to identify and analyse the magnetization precession phenomena corresponding to the observed anticrossing features, a discussion in terms of magnetostatic theory is useful [24-27]. The MSMs resonant frequencies $f$ (and dispersion relation of spin wave modes, generally) in spheroids inserted in MW cavity working at frequency $\omega_c/2\pi$ can be derived from characteristic equation in terms of associated Legendre functions $P_n^m(f, H_0)$:

$$n + 1 + \xi_0 \frac{P_n^{m\prime}(f,H_0)}{P_n^m(f,H_0)} \pm m \frac{\Gamma f M_S}{\Gamma^2 H_i^2 - f^2} = 0, \quad (3)$$

where $H_i = H_0 - M_S/3$, $\xi_0 = \left(1 + \frac{H_i}{M_S} - \frac{f^2}{\Gamma^2 M_S H_i}\right)^{1/2}$ and $P_n^{m\prime}(f, H_0) = \frac{dP_n^m(f,H_0)}{d\xi_0}$.

The index $n \in \mathbb{N}$ labels localization of MSM on the surface with respect to the external magnetostatic field $\vec{H}_0$, while $m$ ($|m| \leq n$) refers to the angular momentum.

The MSMs relative to indexes $n$ and $m$ are then labelled as $(n, m)$ and are grouped in *families* as a function of the value $n - |m|$. Generally the resonant frequencies of MSM are dispersive, except the mode families $n - |m| = 0$ and 1. In these cases, the **Eq. 3** assumes the simplified forms:

$$\frac{f}{\Gamma M_S} - \frac{H_{0,mm}}{M_S} + \frac{1}{3} = \frac{m}{2m+1} \quad (n = m), \tag{4}$$

$$\frac{f}{\Gamma M_S} - \frac{H_{0,m(m+1)}}{M_S} + \frac{1}{3} = \frac{m}{2m+3} \quad (n = m + 1). \tag{5}$$

The **Eq. 4** with fixed $n = m = 1$ gives $f = \frac{\omega_{FMR}}{2\pi} = \Gamma H_{0,11}$, corresponding to uniform magnetization precession.

If the FM sphere is immersed in a confined magnetic field oscillating at frequency $\omega_c/2\pi$ and a strong coupling regime is reached at $H_{0,mn}$, the resonance frequencies of the two subsystems must match. By imposing this condition it is possible to determine the indexes $m$ and $n$ of the MSMs associated to each anticrossing. Thus, after extrapolating the $H_{0,mm}$ values associated to the various anticrossings observed at $f \approx \omega_{101}/2\pi = 8.405$ GHz for the map at TE[101] reported in **Fig. 2.a** [28], a preliminary identification of $(m, m)$ MSM can be carried out. Accordingly to **Eq. 4**, by using an approximate values for $\Gamma$ (also estimated from $H_{0,11} = \frac{\omega_{FMR}}{2\pi\Gamma}$) and of $M_S$ (0.178 T in literature) [24], we exploit the discrete nature of the indexes to facilitate association. Moreover, analysing the $(m, m)$ MSM, the spacing of nine consecutive splittings as a function of the external field is observed to reduce, moving far from FMR condition in **Fig. 2**. Subsequently, the trend of the magnetostatic field $H_{0,mm}$ corresponding to $(m, m)$ MSM anticrossing conditions (**Fig. 4.a**) is exploited to evaluate the saturation magnetization $M_S$, as a fitting parameter for **Eq. 4** [29].

However, the identification of the involved MSM following this procedure is not immediate and hinder further analysis from **Fig. 2**. Starting from this observation and the previously discussed theory, for a more straightforward identification of the different MSM, here we propose as an ansatz to rearrange data in order to plot the resonator signal as a function of cavity frequency and $[H_{0,mm} - (\frac{\omega_c}{2\pi\Gamma} - M_S/6)]^{-1}$ (**Fig. 4.b**). The reason is that (according to mathematical manipulations of **Eq. 4 and 5**), this is expected to bring to equally spaced features for each family when $n - |m| = 0$ or 1, as it is demonstrated from results reported in **Fig. 4.b**. In this frame, the identification of MSM of $(m, m)$ and $(m + 1, m)$ families at $\omega_{101}/2\pi$ is more direct. As a further improvement, in **Fig. 4.c** the signal is plotted as a function of $-1/2 + M_S/4 \left[H_{0,mm} - (\frac{\omega_c}{2\pi\Gamma} - M_S/6)\right]^{-1}$, which is obtained by a rearrangement of **Eq. 4**. It results that the local minima in oscillations of cavity amplitude are shown on an *x*-axis now indicative of the mode $(m, m)$ index, clearly visible until the 9[th] excitation. Notably, in this procedure, even if for $M_S$ a value from literature is employed, anyhow the discrete nature of the indexes would allow a simple association of the mode.

For $(m + 1, m)$ MSM identification, the magnetic field axis should be modified following **Eq. 5**. In this case, the main differences with the previous calculation are the presence of -3/2 as coefficient and a scaling of 3/4 instead of -1/2 and 1/4, respectively. Apparently, several features corresponding to modes (4+3k, 3+3k), k ∈ ℕ, are detected up to $m$ equal to 27. However, the most pronounced absorptions are limited to a few recognizable modes, which are degenerate with respect to other $(m, m)$ ones, thus these features could be also ascribed to the latters. Only smaller $(m + 1, m)$ MSM signatures can be distinguished separately in **Fig. 2.a** (see red arrows), as the (5, 4) and (6, 5) modes. As a result, we conclude that the $(m + 1, m)$ MSM family is less coupled to the cavity.

In **Table 1**, the identified *(m, m)* MSMs observed for the YIG sphere within the cavity for the first transverse electric mode are reported together with their corresponding magnetostatic fields, coupling parameters and photonic/magnonic dampings. Cooperativity *C* among FMR and TE$_{101}$ is at least one order greater than values at other frequency splittings. Moreover, the photonic and magnonic losses were estimated to be comparable for all the different MSM modes.

| $H_0$ (T) | MSM - TE$_{101}$ | $\gamma/2\pi$ (MHZ) | $\kappa/2\pi$ (MHZ) | $g/2\pi$ (MHZ) | C |
|---|---|---|---|---|---|
| 0.2718 | (9, 9) | 3.9 | 4.0 | 4.1 | 1.1 |
| 0.2722 | (8, 8) | 2.9 | 4.2 | 4.7 | 1.8 |
| 0.2728 | (7, 7) | 3.1 | 3.3 | 5.8 | 3.3 |
| 0.2734 | (6, 6) | 3.1 | 3.1 | 7.1 | 5.2 |
| 0.2746 | (5, 5) | 3 | 3.1 | 8.7 | 8.1 |
| 0.2760 | (4, 4) | 3.1 | 2.8 | 10.9 | 13.6 |
| 0.2780 | (3, 3) | 2.5 | 2.5 | 14.4 | 33.2 |
| 0.2826 | (2, 2) | 1.6 | 2.3 | 25.1 | 171.2 |
| 0.2882 | (5, 4) | 2.0 | 4.0 | 8.0 | 8.0 |
| 0.2941 | (1, 1)$_{FMR}$ | 2.5 | 3.2 | 53.5 | 1431.1 |

***Table 1.*** *Summary of obtained parameters for TE$_{101}$ mode.*

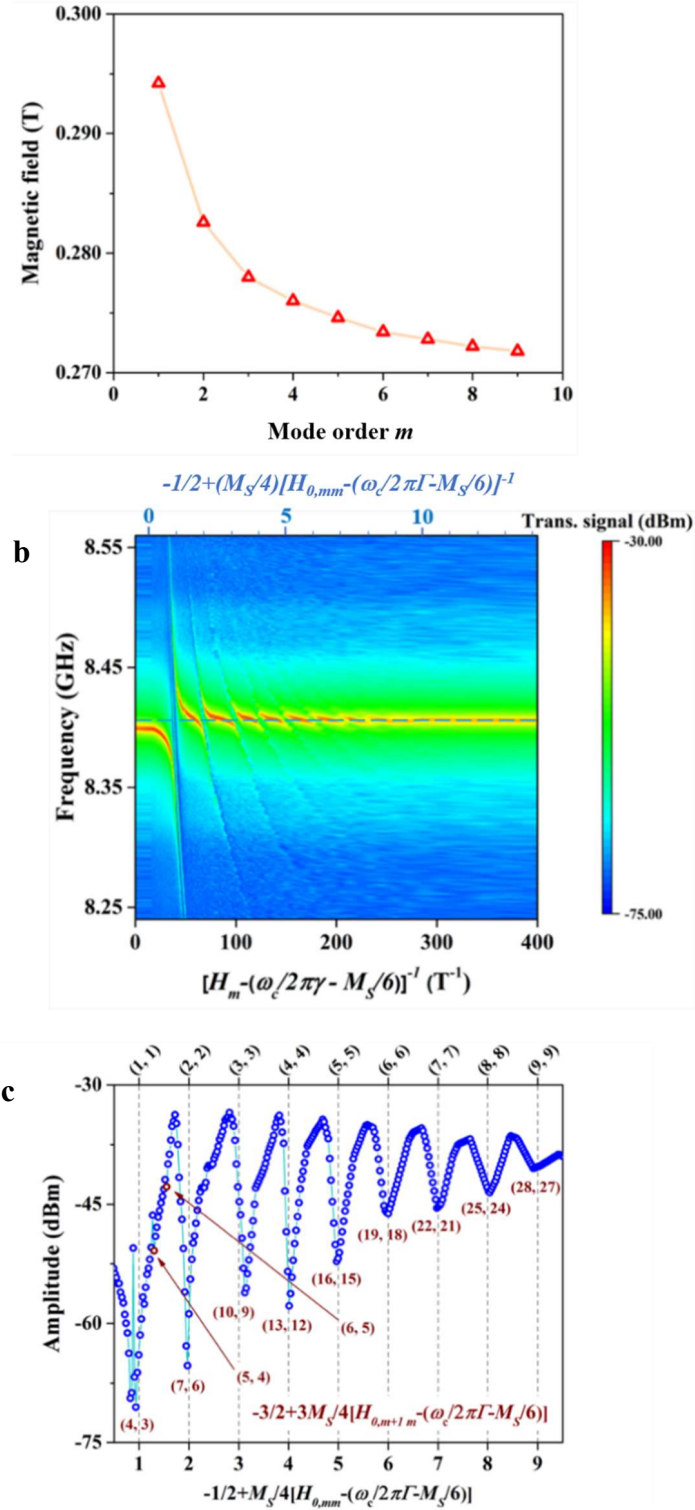

*Figure 4. (a)* Trend of the magnetic fields at resonance conditions as a function of (m, m) MSM order at $TE_{101}$. *(b)* 2D map as a function of frequency near $TE_{101}$ and the introduced functional variable. In this frame, MSMs appear equally spaced. *(c)* At 8.405 GHz, (m, m) modes are recognized up to m = 9. On the other hand, (m+1, m) excitations are not all visible (only the one corresponding to indexes reported in red and in large part being degenerate with (m, m) modes).

*TIME-RESOLVED MEASUREMENTS*

For further insight in the strong coupling regime, we investigate also the time evolution of the strongly coupled system, applying a pulsed 3 μs signal, modulated at the cavity resonance frequency. The signal is collected by an ultrafast digital sampling oscilloscope (Tektronix DSA8300). The input amplitude was set at 10 dBm. The digital sampling oscilloscope is set in order to acquire the envelope of the MW signal by sampling every 125 ps. The spin ensemble dynamics heavily influences the signal amplitude, and both rise and fall times of cavity signal. In **Fig. 5a**, the cavity response at $TE_{101}$ is shown: for lower values of the magnetostatic field, the signal is weakly modified, then it significantly grows for all the pulse duration, with a sensitive increase of cavity relaxation timescale. Successively, by a further increasing of the field, the signal drastically reduces and then increases again and this phenomenon is repeated many times in correspondence of the anticrossing fields. The proposed ansatz was employed to facilitate analysis also of this set of (time-resolved) measurements leading to **Fig. 5.b**, where MSMs up to (9, 9) are visible as increased absorptions [30]. The time scan relative to the coupling among the MW cavity $TE_{101}$ and the (1,1) magnetostatic mode is shown in **Fig. 5.c**. Periodic fluctuations of cavity signal amplitude during charging and relaxation of the system are clearly visible. These oscillations are a further indication of on-resonance hybridization among magnons and photons, which is reached when $\omega_c/2\pi = f_{magnon}$, resulting in a Rabi splitting in two peaks at energies $\hbar/2\pi\,(\omega_c \pm g_i)$. The energy stored inside the cavity decays in an exponential manner, but with periodic oscillations among the two states, demonstrating coherent exchange of energy among photon and magnon modes.

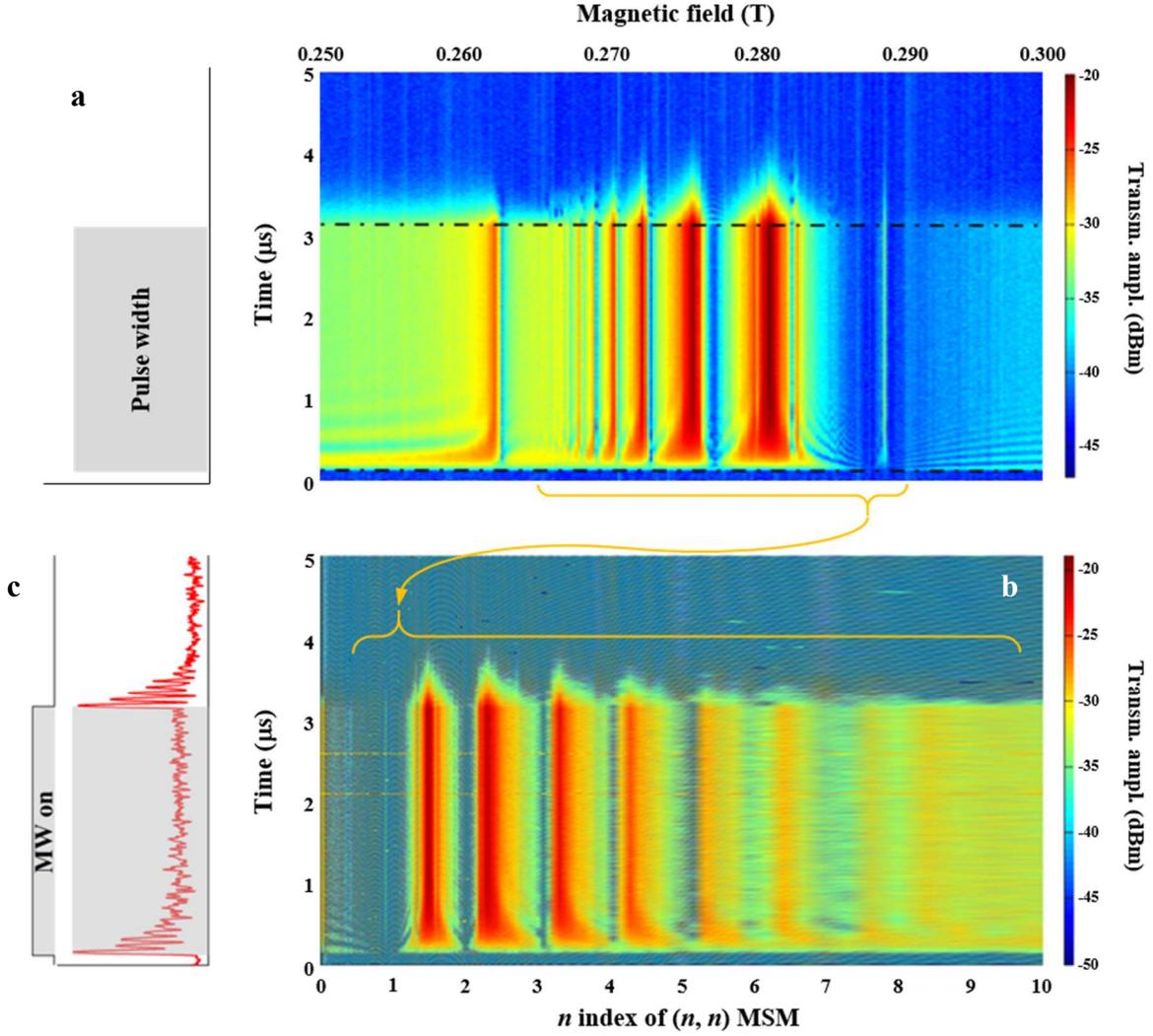

*Figure 5. (a)* Transmission of a rectangular 3 μs microwave pulse at $TE_{101}$ through the cavity versus time and magnetic field. *(b)* time-resolved 2D map with x axis rescaled as a function of MSM index. *(c)* Dynamics at resonance corresponding to strong coupling between $TE_{101}$ cavity mode and (1, 1) MSM.

## CONCLUSION

The strong interaction regime between MSM modes excited in an YIG sphere and photonic modes in a 3D cavity resonator was investigated at RT. A rich spectrum was observed with several anti-crossing features due to coupling to the fundamental FMR mode as well as additional magnetostatic modes. As an ansatz for enabling a simple identification of the various MSMs, we proposed a rescaling procedure in order to plot the resonator signal as a function of cavity frequency and a novel functional variable, just from a mathematical rearrangement of the eigenvalue equations for the MSMs in a spheroids (for both the $(m,m)$ and $(m+1,m)$ families)). This procedure was applied to both the frequency-dependent and time-resolved scans in magnetic field and allowed us to recognize the $(m,m)$ modes as the ones more coupled to the cavity. Magnetostatic fields, coupling parameters, photonic/magnonic dampings and cooperativity $C$ were evaluated for all modes up to the (9,9) one.

Rabi oscillations were clearly visible in time scans, demonstrating (for the first time at room temperature) coherent exchange of energy among photons and the involved magnons modes. The easier identification of magnetostatic modes can be exploited to investigate, control and compare many-levels hybrid systems in cavity- and opto-magnonics research [31-33].

*ACKNOWLEDGEMENTS*

This work was financially supported by the MIUR-PRIN Project (prot.2012EFSHK4) and Regione Puglia NABIDIT – NANOBIOTECNOLOGIE e SVILUPPO PER TERAPIE INNOVATIVE Project (F31D08000050007), L.M. was partially supported by the INFN-IS MMNLP.

*REFERENCES*

1. Dudin, Y. and A. Kuzmich, *Strongly interacting Rydberg excitations of a cold atomic gas.* Science, 2012. **336**(6083): p. 887-889.
2. Putz, S., et al., *Protecting a spin ensemble against decoherence in the strong-coupling regime of cavity QED.* Nature Physics, 2014. **10**(10): p. 720.
3. Niemczyk, T., et al., *Circuit quantum electrodynamics in the ultrastrong-coupling regime.* Nature Physics, 2010. **6**(10): p. 772.
4. Bienfait, A., et al., *Reaching the quantum limit of sensitivity in electron spin resonance.* Nature nanotechnology, 2016. **11**(3): p. 253.
5. Bozhko, D.A., et al., *Supercurrent in a room-temperature Bose–Einstein magnon condensate.* Nature Physics, 2016. **12**(11): p. 1057.
6. A. Leo, S.R., A. G. Monteduro and G. Maruccio, *Strong Coupling in Cavity Magnonics*, in *Three-Dimensional Magnonics: Layered Micro-and Nanostructures*. 2019, Jenny Stanford Publishing: New York.
7. Kittel, C., *On the theory of ferromagnetic resonance absorption.* Physical Review, 1948. **73**(2): p. 155.
8. Zhang, X., et al., *Strongly coupled magnons and cavity microwave photons.* Physical review letters, 2014. **113**(15): p. 156401.
9. Tabuchi, Y., et al., *Hybridizing ferromagnetic magnons and microwave photons in the quantum limit.* Physical review letters, 2014. **113**(8): p. 083603.
10. Wang, Y.-P., et al., *Bistability of Cavity Magnon Polaritons.* Physical Review Letters, 2018. **120**(5): p. 057202.
11. Zhang, X., et al., *Magnon dark modes and gradient memory.* Nature communications, 2015. **6**.
12. Goryachev, M., et al., *High-cooperativity cavity QED with magnons at microwave frequencies.* Physical Review Applied, 2014. **2**(5): p. 054002.
13. Lambert, N., J. Haigh, and A. Ferguson, *Identification of spin wave modes in yttrium iron garnet strongly coupled to a co-axial cavity.* Journal of Applied Physics, 2015. **117**(5): p. 053910.
14. Zhang, X., et al., *Superstrong coupling of thin film magnetostatic waves with microwave cavity.* Journal of Applied Physics, 2016. **119**(2): p. 023905.
15. Morris, R., et al., *Strong coupling of magnons in a YIG sphere to photons in a planar superconducting resonator in the quantum limit.* Scientific Reports, 2017. **7**.
16. The single crystal YIG sphere of 1 mm diameter is even used as Standard Reference Material (SRM) for calibration of magnetometers; the certificate of analysis is reported in: https://www-s.nist.gov/srmors/certificates/2853.pdf
17. Cherepanov, V., I. Kolokolov, and V. L'vov, *The saga of YIG: spectra, thermodynamics, interaction and relaxation of magnons in a complex magnet.* Physics reports, 1993. **229**(3): p. 81-144.
18. Fano, U., *Effects of configuration interaction on intensities and phase shifts.* Physical Review, 1961. **124**(6): p. 1866.

19. Kamenetskii, E., G. Vaisman, and R. Shavit, *Fano resonances of microwave structures with embedded magneto-dipolar quantum dots.* Journal of Applied Physics, 2013. **114**(17): p. 173902.
20. Harder, M., et al., *Study of the cavity-magnon-polariton transmission line shape.* Science China Physics, Mechanics & Astronomy, 2016. **59**(11): p. 117511.
21. Walls, D.F. and G.J. Milburn, *Quantum optics.* 2007: Springer Science & Business Media.
22. Agarwal, G., *Vacuum-field Rabi splittings in microwave absorption by Rydberg atoms in a cavity.* Physical review letters, 1984. **53**(18): p. 1732.
23. Stancil, D.D. and A. Prabhakar, *Spin waves.* 2009: Springer.
24. Röschmann, P. and H. Dötsch, *Properties of magnetostatic modes in ferrimagnetic spheroids.* physica status solidi (b), 1977. **82**(1): p. 11-57.
25. Fletcher, P. and R. Bell, *Ferrimagnetic resonance modes in spheres.* Journal of Applied Physics, 1959. **30**(5): p. 687-698.
26. White, R.L., *Use of magnetostatic modes as a research tool.* Journal of Applied Physics, 1960. **31**(5): p. S86-S94.
27. Walker, L., *Magnetostatic modes in ferromagnetic resonance.* Physical Review, 1957. **105**(2): p. 390.
28. Slightly shifted from 8.401 GHz due to proximity to strong coupling condition.
29. The estimated value of gyromagnetic ratio is 28.76 GHz/T, while the experimental value of saturatioin magnetization is 0.176 T.
30. Values estimated from time-resolved set of data: gyromagnetic ratio 29.24 GHz/T, saturation magnetization 149.49 mT.
31. Liu, Z.-X., et al., *Phase-mediated magnon chaos-order transition in cavity optomagnonics.* Optics letters, 2019. **44**(3): p. 507-510.
32. Yang, Y., et al., *Control of the magnon-photon level attraction in a planar cavity.* Physical Review Applied, 2019. **11**(5): p. 054023.
33. Bhoi, B., et al., *Abnormal anticrossing effect in photon-magnon coupling.* Physical Review B, 2019. **99**(13): p. 134426.